%% file: article.tex
	\documentclass[preprint,twocolumn,secnumarabic,amssymb,nobibnotes,nofootinbib,10pt]{revtex4-1}
	\usepackage[english]{babel}
	\usepackage{graphicx}

	\usepackage{amssymb}
	\usepackage{amsmath,mathtools}
	\usepackage{amsfonts}
	\usepackage[usenames,dvipsnames]{color}
	\usepackage{hyperref}
	\usepackage{rotating}
	\usepackage{slashed}
	\usepackage{transparent}
	\newcommand{\kBT}[0]{k_\mathrm{B}T}

	\DeclareMathAlphabet{\mathbb}{U}{bbold}{m}{n}

\begin{document}
	\title{Fundamental measure theory for the electric double layer: implications for blue-energy harvesting and water desalination}
	\author{Andreas H\"{a}rtel, Mathijs Janssen, Sela Samin and Ren\'{e} van Roij}
	\address{Institute for Theoretical Physics, 
	Center for Extreme Matter and Emergent Phenomena, 
	Utrecht University, Leuvenlaan 4, 3584 CE Utrecht, The Netherlands}
	%
	%

	\begin{abstract}
	Capacitive mixing (CAPMIX) and capacitive deionization (CDI) are promising candidates for 
	harvesting clean, renewable energy and for the energy efficient production of potable water, 
	respectively. Both CAPMIX and CDI involve water-immersed porous carbon (supercapacitors) electrodes 
	at voltages of the order of hundreds of millivolts, such that counter-ionic packing is important 
	for the electric double layer (EDL) which forms near the surfaces of these porous materials. Thus, 
	we propose a density functional theory (DFT) to model the EDL, 
	where the White-Bear mark II fundamental measure theory functional is combined with a mean-field 
	Coulombic and a mean spherical approximation-type correction to describe the interplay between dense packing and electrostatics, 
	in good agreement with molecular dynamics simulations. 
	We discuss the concentration-dependent potential rise due to changes in the chemical potential in 
	capacitors in the context of an over-ideal theoretical description and its impact on energy 
	harvesting and water desalination. 
	Compared to less elaborate mean-field models our DFT calculations reveal a higher work output for 
	blue-energy cycles and a higher energy demand for desalination cycles. 

	\end{abstract}
	\maketitle

	\section{Introduction}

	{\it 

	}

	Two of the most stringent challenges mankind faces 
	in the coming century are the ever-growing worldwide 
	demands for energy and fresh water. 
	The shortages in these resources cannot be seen separately since the 
	generation of fresh water with current techniques is an energy 
	consuming process. However, either energy or fresh water often 
	is available regionally. 
	Thus, apart from alternative sources like 
	offshore ground water \cite{post_nature504_2013} or 
	solar and wind power, 
	it is of interest to consider low-cost processes which generate energy 'using' fresh water 
	and the other way round. 

	It has been known for a long time that the irreversible mixing of two distinct fluids 
	is a spontaneous process which rises the entropy and dissipates free energy. 
	In the case of mixing fresh and salty water that occurs when rivers flow into 
	the sea, this Gibbs free energy of mixing amounts to (the order of) $2$kJ per 
	liter of river water, the equivalent of a 200m 
	waterfall \cite{norman_science186_1974}. Considering the worldwide amount of 
	water flowing into the sea, the harvest of this energy could account for a 
	few percent of the global energy needs. 
	Furthermore, the reverse process de-mixes salt water 
	into brine and fresh water and 
	requires at least the same amount of energy as its corresponding blue-energy process 
	harvests. 

	The Gibbs free energy of mixing is harvested in so-called blue engines 
	which often rely on membranes to selectively intercept some of the 
	involved particles \cite{logan2012membrane}, either cations, anions, or water. 
	Test factories have been built, based either on water-permeable 
	membranes in the case of pressure retarded osmosis \cite{gerstandt2008membrane, lin2014thermodynamic} 
	or ion-selective membranes in the case of reverse electro dialysis 
	\cite{post2007salinity}. 
	The downside of all these techniques however 
	is that membranes are costly and tend to foul upon exposure to the large 
	amounts of water flowing through them. 
	%

	A promising alternative to the techniques mentioned above has recently 
	been offered by Brogioli \cite{brogioli2009extracting}. He proposed the 
	use of porous carbon electrodes to selectively mix the liquids
	in a four-stroke charging-desalination-discharging-resalination cycle, 
	very much in the spirit of a Stirling heat engine that performs an 
	expansion-cooling-compression-reheating cycle. Within this capacitive 
	mixing (CAPMIX) process \cite{rica_entropy15_2013}, the charging of the 
	electrodes takes place while immersed in salty water, whereas they are 
	discharged in fresh water. The de- and resalination steps are 
	performed by flushing the electrodes with 
	fresh and sea water, respectively, which enlarges or compresses the 
	electric double layer (EDL). 
	Such cycles have a net positive work output, because the expansion/compression 
	of the EDL changes the capacitance and thereby the potential (at fixed charge 
	on the electrode). More specifically, salty water 
	more effectively screens the charge on the electrode surfaces such that 
	charging steps are performed at lower potential than discharging steps. 
	In the reverse process of capacitive desalination (CDI) 
	energy is consumed to lower the salinity of a finite volume (say a bucket) of water, 
	while salt ions are released into the sea or a reservoir 
	(the performance of this process has been studied in \cite{porada_ees6_2013}). 

	Interestingly, all 
	devices discussed have a broader applicability than just the mixing of sea- 
	and river water since they do not put severe restrictions on the type of 
	solutions that are mixed. 
	For instance, energy can also be harvested from mixing 
	CO$_{2}$ with air \cite{hamelers2013harvesting}, because CO$_{2}$ 
	produces charged carbonic acids when it is flushed through water. 
	These acid particles can subsequently play the same role in a CAPMIX process as 
	the salt ions in water in a conventional blue engine cycle.

	In contrast to membranes, porous carbons couple a huge internal surface area of the order 
	of a square km per kg to long lifetimes and low manufacturing costs 
	\cite{simon2008materials,zhu_science332_2011}. For this reason, 
	they can for instance also be used in ionic liquid-filled (super)capacitors to store large amounts 
	of energy. 
	In fact, due to the safety of ionic liquids and high (dis)charging rates, supercapacitors 
	are already used in buses and planes. 
	The porous-material properties are highly determined by the distribution of pore-sizes 
	\cite{lastoskie_jpc97_1993}, 
	where the physics inside a pore depends on its shape 
	\cite{huang_cej14_2008,merlet_jpcl4_2013} and size 
	\cite{jiang_nl11_2011,kondrat_ees5_2012,jiang_jpcl3_2012}. 
	%
	%
	Thus, tuning the porous structure of the material \cite{tran_jps235_2013} 
	helps to optimize the storage of energy within supercapacitors. 
	These findings motivate us to investigate whether similar phenomena can be observed 
	in the aqueous electrolyte-filled supercapacitors of blue engines and desalination devices. 
	As a starting point we focus on a quasi-static description leaving 
	dynamic effects associated with optimal power \cite{marino_jcis436_2014} 
	for later studies. 

	The (macroscopic) behaviour of the materials and devices discussed above is 
	determined by the EDL which forms in the (microscopic) region close to the 
	electrodes' surface. A lot of research has been performed on its theoretical 
	modeling, dating back to Gouy-Chapman's analytical solution of 
	the Poisson-Boltzmann equation for point-like ions in 
	contact with a charged planar wall \cite{gouy1910constitution,Chapman1913}. 
	However, this simple theory fails to capture the rich variety 
	of phenomena as observed in more advanced models that do take into account
	the steric repulsion between ions in the EDL, e.g. 
	simulation studies 
	\cite{torrie_jcp73_1980,fedorov_jpclb112_2008,merlet2012molecular}, phenomenological extensions 
	\cite{kornyshev2007double,Bazant:2011aa}, or bottom-up density functional approaches 
	\cite{wu_aiche52_2006,biben_pre57_1998,
	mieryteran_jcp92_1990,gillespie_pre68_2003,antypov_pre71_2005,
	jiang_cpl504_2011,forsman_jpcb115_2011,wang_jpcm23_2011,frydel_jcp137_2012,henderson_jpcb116_2012}, 
	leading to phenomena like 
	a bell-shape capacitance at high salt concentrations \cite{kornyshev2007double}, 
	oscillations in capacitance \cite{fedorov_jpclb112_2008}, 
	and an anomalous capacitance increase in sub-nanometer pores \cite{merlet2012molecular}. 
	%
	%
	%
	%
	%
	%
	%
	%
	%
	%
	In the case of (classical) density functional theory \cite{evans_ap28_1979}, 
	the hard-sphere repulsion is included from first principles. 
	Beyond the restricted primitive model of equally charged hard spheres 
	in contact with a charged hard wall, further physics has been addressed, 
	for example by
	a more realistic treatment of water \cite{jeanmairet_jpcl4_2013}, 
	the polarization of particles \cite{jiang_jpcl4_2013}, 
	or the asymmetry of ions \cite{georgi_jec649_2010,han_jpcm26_2014,breitsprecher_jpcm26_2014}. 

	In this work, we study blue-energy and desalination 
	devices within the restricted primitive model using a modern density functional 
	theory (DFT) approach. To construct our functional, we follow the work of 
	Mier-y-Teran et al. \cite{mieryteran_jcp92_1990}, including pure Coulombic and 
	hard-sphere correlations as well as residual correlations from the mean spherical approximation (MSA). 
	Thereby, we avoid the use of an arbitrary weight function 
	that enters in the recently proposed weighted correlation approach FMT/WCA-$k^2$ 
	\cite{wang_jpcm23_2011}, which has been used in a study of the 
	electric double layer in slit-like pores \cite{pizio_cmp17_2014}. 
	In contrast to previous work 
	\cite{mieryteran_jcp92_1990,tang_MolPhys71_1990,yu_cjce12_2004,yu_jcp_2004,
	jiang_cpl504_2011,wang_jpcm23_2011,
	henderson_jpcb116_2012,jiang_jpcl3_2012,jiang_jpcl4_2013,pizio_cmp17_2014}, 
	we describe the pure hard-sphere interactions using the White Bear mark II 
	approach \cite{hansen-goos_jpcm18_2006} within fundamental measure theory 
	\cite{rosenfeld_prl63_1989,roth_jpcm22_2010}. To our knowledge, 
	this approach describes hard spheres within DFT most accurately and even 
	predicts the crystalline phase and its coexistence with the fluid phase 
	quantitatively \cite{oettel_pre82_2010,haertel_prl108_2012}. 
	Our research focusses on the fundamental understanding of the 
	cyclic CAPMIX and CDI processes, in particular on the molecular 
	structure of the densely-packed electrolyte confined in the pores of the 
	charged nanoporous electrode. 
	Finally, we discuss optimal cycle characteristics, taking previous 
	thermodynamical consideration into account \cite{boon2011blue,Roij:2012aa}. 

	\section{Set-up}

	\begin{figure}[t]
	\centering
	\def\svgwidth{0.5\textwidth}{ 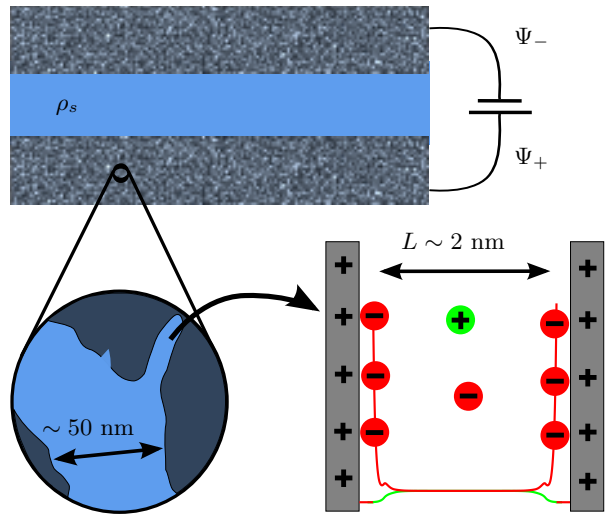}
	\caption{\label{fig:set_up}Sketch of the supercapacitor as used in 
	blue engines and desalination devices. 
	Two porous carbon electrodes are set to the potentials $\Psi_{+}$ and $\Psi_{-}$ 
	and are immersed in a monovalent ion solution with salt concentration $\rho_{\rm s}$. 
	The porous material contains pores of several sizes. The small pores are 
	modeled as parallel plate capacitors of plate separation or pore size $L$. 
	}
	\end{figure} 

	A so-called {\it blue engine} consists of a supercapacitor 
	which can be connected to sea and river reservoirs and performs a sequence of charging 
	and discharging steps to harvest {\it blue energy}. 
	A {\it desalination device} also consists of a supercapacitor, but it is connected to a 
	finite-sized bucket which is desalinated and subsequently refilled (after securing the fresh 
	water) with salty water from 
	a sea reservoir in a cyclic fashion. 

	In our work, the supercapacitor consists of two porous carbon electrodes at potentials 
	$\Psi_{+}$ and $\Psi_{-}$ as sketched in figure \ref{fig:set_up}. 
	Here, a typical porous electrode consists of a broad distribution of pore sizes 
	which are classified as macropores ($>50$nm), mesopores ($2-50$nm), and 
	micropores ($<2$nm) \cite{rouquerol_pac66_1994}. 
	The macropores act as transport channels and are globally charge neutral, 
	whereas the micropores have most impact on the storage of charges, in particular 
	due to overlapping double layers. 
	We model the electrode to consist solely of equally sized pores, where we 
	choose either micropores of a typical size $L=2$nm within such electrodes or 
	mesopores of $L=8$nm, a width were packing constraints are less severe. 
	The pores in one electrode have a total volume $V_{\rm el}=L\times A$ 
	and are described by a parallel plate capacitor of plate area $A$ and separation $L$; 
	we call $L$ the pore size. The capacitor plates are 
	located at $z=0$ and $z=L$ and are kept at the same 
	potential as the macroscopic electrode they belong to. Accordingly, 
	they carry like-charge densities $e\sigma$, 
	resulting in a total electrode charge $Q=2A e\sigma$, 
	where we introduced the elementary unit charge $e$. 
	Furthermore, we neglect size effects of the parallel plate capacitor by taking the 
	limit $A\rightarrow\infty$ such that the system is described 
	solely by one spatial coordinate $z$, perpendicular to the plates. 
	For symmetry reasons we only need to model the macroscopic anode of our blue engine 
	with a pore volume $V_{\rm el}$ and a pore size $L$. The cathode is just an 
	oppositely charged mirror image of the anode. 

	%
	%
	%
	To describe the ions in water we apply the restricted primitive 
	model, where a binary mixture of oppositely charged hard spheres is confined 
	between two planar charged hard walls. Accordingly, 
	the cations and anions are modeled by monodisperse charged hard spheres of radius 
	$R=R_{+}=R_{-}=0.34$nm (as used in \cite{boon2011blue,Janssen:2014aa})
	that carry the charges $Z_{+}e$ and $Z_{-}e$, 
	respectively, and already contain the surrounding hydration shell of polarized 
	solvent molecules for each ion. In this work, the ions of interest are Na$^{+}$ 
	and Cl$^{-}$ ions of valencies $Z_{+}=1$ and $Z_{-}=-1$. As in previous 
	work \cite{wang_jcp135_2011,luksic_molphys110_2012}, 
	the remaining homogeneous dielectric solvent is characterized by a relative 
	dielectric constant $\varepsilon=78.54$ at a temperature $T=298.15$K. 
	For this setting, the Bjerrum length 
	$\lambda_{\rm B}=e^2/(4\pi\varepsilon_0\varepsilon\kBT)$ with the vacuum permittivity 
	$\varepsilon_0$ and the Boltzmann constant $k_{\rm B}$ 
	amounts to $\lambda_{\rm B}=0.714$nm. The resulting Debye screening length 
	$\kappa^{-1}$ with $\kappa^2=4\pi\lambda_{\rm B}(Z_{+}^2+Z_{-}^2)\rho_{\rm s}$, 
	ranges from $\kappa^{-1}=0.43$nm to $\kappa^{-1}=2.15$nm for
	bulk salt concentrations between $\rho_{\rm s}=0.5$M (sea water) and $\rho_{\rm s}=0.02$M 
	(river water), respectively. 
	Now, the particle interaction between two ions $i$ and $j$ at a distance $r$ reads 
	\begin{equation}
	\Phi_{ij}(r) 
	= 
	\begin{cases}
	\displaystyle{\infty } & r<2R; \\
	\displaystyle{\kBT \lambda_{\rm B} \frac{Z_i Z_j}{r}} & r\geq 2R.
	\end{cases}  
	\end{equation}
	Similarly, the steric interaction between the pore walls 
	and a cation or anion at position $z$ is 
	\begin{equation}
	V_{\pm}^{\rm ext}(z) 
	= \kBT
	\begin{cases}
	\displaystyle{\infty } & (z<R_{\pm}) \, \mathrm{~or~} \, (L-z<R_{\pm}); \\
	\displaystyle{0} & \mathrm{otherwise}.
	\end{cases}  \label{eq:external_field_wall}
	\end{equation}
	Here, the steric repulsion  between the hydrated ions and the wall naturally leads to 
	a so-called Stern layer of the same thickness as the hydrated-ion radius $R$. 
	Within this Stern layer, the concentration of charged particles remains zero, while 
	we assume no change of the dielectric constant to avoid additional polarization effects. 



	Now, we define the overall unit-charge density as 
	\begin{equation}
	q\big(z;[\rho_{+},\rho_{-},\sigma]\big) = Z_{+}\rho_{+}(z) + Z_{-}\rho_{-}(z) + 
	\sigma \delta(z)+\sigma \delta(z-L) , 
	\label{eq:overall_charge_density}
	\end{equation}
	containing contributions from cations and anions via their one-particle number 
	densities $\rho_{\pm}$ and from the surface charge density $e\sigma$ 
	localized on the electrodes at $z=0$ and $z=L$. 

	\section{Theory}


	We use density functional theory (DFT) \cite{evans_ap28_1979,tarazona_inbook_2008} 
	to describe the system of densely-packed 
	ions and charged walls within each of the macroscopic electrodes. 
	Within this theoretical framework, the grand 
	potential $\Omega(T,V,\mu_{+},\mu_{-},\Psi)$ of the equilibrated system can be written as a functional 
	$\Omega(T,V,\mu_{+},\mu_{-},\Psi;[\rho_{+},\rho_{-},\sigma])$ of the one-particle number densities 
	$\rho_{\pm}$ of the cations and anions and of the wall unit-charge density $\sigma$ 
	for a temperature $T$, a volume $V$, ionic chemical potentials $\mu_{\pm}$ 
	and an electrostatic wall-potential $\Psi=\Psi_{+}=-\Psi_{-}$. 
	The grand potential naturally follows from the intrinsic Helmholtz free energy $F(T,V,N_{+},N_{-},Q)$ 
	via a generalized Legendre transform
	\begin{eqnarray}
	&&\Omega\big(T,V,\mu_{+},\mu_{-},\Psi;[\rho_{+},\rho_{-},\sigma]\big) \nonumber \\
	&&\hspace{0.5cm}= F\big(T,V;[\rho_{+},\rho_{-},\sigma]\big) \\
	& &\hspace{0.9cm}
	- 2 e \Psi A \sigma 
	+ \sum_{i=\pm} A\int_{0}^{L}\Big( V_{i}^{\rm ext}(z) - \mu_{i} \Big) \rho_{i}(z) dz , \nonumber
	\end{eqnarray}
	where the number density profiles $\rho_{\pm}$ and the unit-charge density profile $\sigma$ 
	represent the extensive particle numbers $N_{\pm}=A\int_0^L \rho_{\pm}(z) dz$ and 
	electrode charge $Q=2e A\sigma$. 
	In the presence of particle interactions the intrinsic 
	free energy $F=F_{\rm id}+F_{\rm exc}$ is usually split into an ideal gas part 
	$F_{\rm id}$ 
	(see e.g. Ref.~\cite{hansen_book_2006}) 
	and an excess part $F_{\rm exc}$ that 
	contains the (remaining) interactions. 
	Minimized by the equilibrium density and charge profiles, the functional finally yields 
	the Euler-Lagrange equations 
	\begin{eqnarray}
	\mu_{\pm} &=& V_{\pm}^{\rm ext}(z) 
	 + \left.\frac{\delta F\big(T,V;[\rho_{+},\rho_{-},\sigma]\big)}{\delta\rho_{\pm}(z)}
	\right|_{\rho_{\pm}=\rho_{\pm}^{\rm eq}} , \label{eq:ele1_rho} \\
	\Psi &=& (2 e A)^{-1} \left.\frac{\delta F_{exc}\big(T,V;[\rho_{+},\rho_{-},\sigma]\big)}{\delta\sigma}
	\right|_{\sigma=\sigma^{\rm eq}} . 
	\label{eq:ele1_sigma}
	\end{eqnarray}


As in previous work \cite{mieryteran_jcp92_1990,yu_cjce12_2004}, 
the excess free energy functional 
$F_{\rm exc}=F_{\rm C} + F_{\rm HS} + F_{\rm corr}$ 
in our restricted primitive model contains 
the excess terms $F_{\rm C}$ and $F_{\rm HS}$
due to the pure Coulomb and hard-sphere interactions, as well as 
an excess term $F_{\rm corr}$ which gives rise to residual correlations. 
Here, the excess correlation term $F_{\rm corr}$ follows from the analytically known 
solution of the Ornstein-Zernike equation in the mean spherical 
approximation (MSA) closure in the bulk \cite{waisman_jcp52_1970}. 
Furthermore, 
the mean-field Coulomb term $F_{\rm C}$ typically 
defines the (dimensionless) electrostatic potential $\phi(z)=\tfrac{e}{\kBT}\psi(z)$ 
via the formal solution of the Poisson equation. 
This electrostatic potential connects to the electrostatic wall potential $\Psi$ via 
$\Psi=\psi(0)=\psi(L)$ and allows to extract 
the unit-charge density $\sigma$ 
using Gauss' law. 
New in our description, 
the hard-sphere interactions are modeled within fundamental measure theory (FMT) 
via the White-Bear mark II functional \cite{hansen-goos_jpcm18_2006} in its tensor 
version \cite{oettel_pre82_2010}. This functional 
successfully predicts not only the freezing transition and correct free energies in 
a hard-sphere system \cite{hansen-goos_jpcm18_2006,oettel_pre82_2010}, but also 
the interfacial tension of a crystal-fluid interface \cite{haertel_prl108_2012}. 
Henceforth, we call this density functional approach {\it FMT-PB-MSA}, 
referring to the three contributions to the excess free energy. For comparisons, 
we also introduce {\it FMT-PB}, where we set the excess term 
$F_{\rm corr}=0$.



In the limit $R\to 0$ of charged point-like ions, 
the excess part $F_{\rm HS}$ of the free energy that describes 
the short-ranged hard-sphere interactions vanishes. Accordingly, the excess 
correlation term $F_{\rm corr}$ also disappears. 
To preserve the Stern layer in this limit, we do not apply $R\to 0$ to the 
external potential $V_{\pm}^{\rm ext}$ in equation~(\ref{eq:external_field_wall}) for this 
{\it PB+S} approach, 
but replace the particle radius $R$ with the thickness $R_{\rm S}$ of the Stern layer. 
We furthermore consider the case without Stern layer by setting $R_{\rm S}=0$, henceforth 
referred to as {\it PB}. 
For comparison, we also consider the so-called modified Poisson-Boltzmann theory ({\it mPB}) 
\cite{kornyshev2007double}, where 
steric ion effects are included using 
a lattice-gas ansatz. 

\subsection{Simulation methods}

To test our theoretical results, we perform molecular dynamics simulations ({\it Sim}) 
of spherical particles with diameters $2R$ and charges $\pm e$ using the ESPResSo 
package \cite{espresso,espresso2}. 
For the short-range part of the pair interaction we use the pseudo hard-sphere potential 
\begin{align}
 u_{ij}=
\begin{cases}
 \frac{50^{50}}{49^{49}} \epsilon\left[\left(\frac { 2R } { r_{ij} }
\right)^ { 50 }
-\left(\frac{2R}{r_{ij}}\right)^{49} \right]+\epsilon, 
& r_{ij}<\frac{50}{49}2R \\
0 , & r_{ij}\ge\frac{50}{49}2R
\end{cases}
\label{eq:phs}
\end{align}
where $r_{ij}$ is the distance between two particles 
$i$ and $j$ and $\epsilon$ is the interaction parameter. This cut-and-shifted 
generalized Lennard-Jones potential is suitable for use in continuous molecular dynamics 
simulations and was found to reproduce structural and thermodynamical data of hard-spheres 
accurately over the whole fluid range \cite{jover2012}. 
The particles are confined between two charged hard walls of distance $L$, 
whereby the short range interaction with the walls is also given by equation~(\ref{eq:phs}). 
The long-ranged Coulomb forces were evaluated using the P$^3$M method \cite{arnold2002} 
with metallic boundary conditions. The unwanted interaction between periodically 
replicated slabs is subtracted using the electrostatic layer correction method \cite{arnold2002}. 

In order to compare with our FMT-PB-MSA results, we perform simulations in the canonical 
ensemble using the ion number densities obtained from the DFT calculations, which also 
sets the charge density on the pore walls. Thereby, the number 
of particles in the pore varies between 3000-6000. 
We find that for large enough pores, the mid-plane bulk ion densities 
$\rho_{\pm}(L/2)$ in all the simulations agree with the FMT-PB-MSA result to within $0.5\%$. 
This indicates that our canonical simulation results agree with the grand-canonical FMT-PB-MSA 
calculations. 
The simulations are initialized from random particle configurations and are equilibrated 
with a Langevin thermostat at a temperature $\kBT/\epsilon=1$. We use a time step of 
$\Delta t = 0.001-0.004$ (in standard simulation units), 
where smaller values are required for larger number 
densities. After an equilibration of the system for at least $2\times10^5$ 
time steps, the system was sampled every $200-800$ time steps for at least $10^6$ time steps. 
The obtained cation- and anion-density profiles are used for the calculation of the 
electrostatic potential profile from
\begin{align}
 \phi(z)=-2\pi \lambda_{\rm B}
 \int_0^z \left[\int_0^{z'} q(z'') {\rm d}z''-\int_{z'}^L q(z''){\rm d}z'' \right] dz' ,
\end{align}
where $q(z)$ is defined in equation~(\ref{eq:overall_charge_density}).
This relation yields the potential in slit systems with Neumann boundary conditions 
up to an arbitrary constant \cite{kovacs2012}. We choose this constant such that the 
midplane potential $\phi(L/2)$ is identical to that obtained from FMT-PB-MSA.

\section{Concentration profiles}

To obtain equilibrium concentration profiles $\rho_{\pm}(z)$, 
we solve the Euler-Lagrange equations~(\ref{eq:ele1_rho}) and 
(\ref{eq:ele1_sigma}) 
on a grid using a Picard iteration scheme \cite{ng_jcp61_1974}. 
Assuming a homogeneous concentration in directions along the pore 
walls, we resolve the pore width $L$ in $z$-direction 
perpendicular to the walls 
with a grid spacing of $L/2000\leq 0.004$nm. With this choice the 
typical length scale of the system (screening length $\kappa^{-1}$ or hard sphere diameter $2R$) 
is resolved with more than $100$ grid points. 
We explore electrode potentials over the whole 
electrochemical window of water, where the limit of $2\Psi=1.2$V represents the 
starting point of electrochemical reactions between the electrode and the 
solvent particles. 

\begin{figure}
\centering
\includegraphics[width=8.5cm]{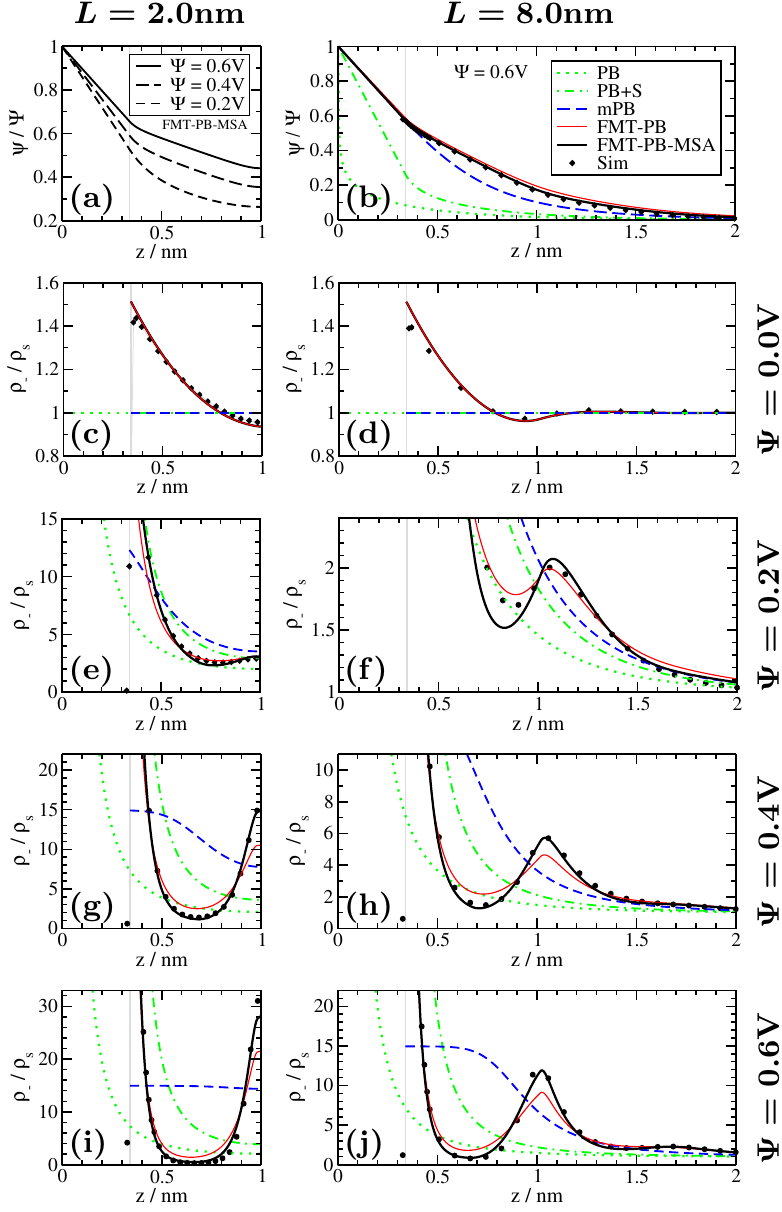}
\caption{\label{fig:conc_profiles} 
(a),(b) Potential profiles $\psi(z)$ and (c)-(j) counterion concentration profiles $\rho_{-}(z)$ 
at an electrode potential $\Psi$ and a bulk concentration $\rho_{\rm s}=0.5$M. 
We show data for two different pore sizes of (left column) $L=2$nm and 
(right column) $L=8$nm as well as for four electrode potentials 
$\Psi$ of (c),(d) $0.0$V, (e),(f) $0.2$V, (g),(h) $0.4$V, and (i),(j) $0.6$V. 
The potential profiles in (a) are solely obtained from FMT-PB-MSA theory, where all 
the remaining plots show data obtained from all theories investigated and our 
molecular dynamics simulations. }
\end{figure} 

In figure \ref{fig:conc_profiles}, we show the resulting potential and concentration profiles 
for the various theories introduced in the 
previous section in comparison with our molecular dynamics simulations. 
In figures \ref{fig:conc_profiles}(a) and (b), we show potential 
profiles which are normalized to unity at the wall for two pores of sizes 
$L=2$nm and $L=8$nm. The corresponding concentration profiles are shown in 
figures \ref{fig:conc_profiles}(c)-(j), where we focus on the region close to the wall. 
Note that 
the profiles within a pore are mirror symmetric with respect to the pore center. 

Figure~\ref{fig:conc_profiles}(a) shows three potential profiles from FMT-PB-MSA 
theory at different electrode potentials $\Psi$. Next to the wall, all of them have 
a linear part due to an ion-free Stern layer which FMT-type theories naturally predict. 
At high potentials, even a second Stern layer can be observed, indicating a 
second region free of charges in between the first and second layer of densely packed 
counterions next to the wall. Such layering effects 
have been observed e.g. in \cite{fedorov_jpclb112_2008,oguz_jpcm21_2009} 
and they are clearly visible in figure \ref{fig:conc_profiles}(i). Thus, high 
enough potentials repel the coions completely and lead to an increasing midplane potential 
$\psi(L/2)$ for increasing electrode potentials $\Psi$. For larger pores this effect 
diminishes, which can be observed in figure \ref{fig:conc_profiles}(b) for the mesopore 
of size $L=8$nm, where the center of the pore behaves similar to the bulk. 
Therefore, the mesopores play a smaller role in the salt uptake, because 
an excess number of ions is only contained in 
the electric double layer. 

In figures \ref{fig:conc_profiles}(c) and (d), we start our analysis of the concentration 
profiles with a vanishing external potential $\Psi=0$. 
In this case the pore walls are uncharged and only the FMT-type theories 
show inhomogeneous concentration profiles due to built-in steric interactions. 
Since the ions are monodisperse, both co- and counterions have the same 
concentration profiles and, counterintuitively, 
lead to a negative net salt adsorption due to the naturally predicted ion-free 
Stern layer. This effect 
has already been observed for pure hard-sphere systems \cite{deb_jcp134_2011}. 
For the non-vanishing potentials, PB-type models cannot capture these packing effects 
since they are based on an ideal gas description. Especially 
mPB theory 
underestimates the concentrations close to the electrodes and, even more, predicts 
a plateau for higher packings \cite{frydel_jcp137_2012} as shown in 
figures~\ref{fig:conc_profiles}(g), (i) and (j). 
Here, the almost flat concentration profile in figure \ref{fig:conc_profiles}(i) 
signifies an almost completely filled pore. 

Finally, our computer simulations show excellent agreement with FMT-PB-MSA theory. 
As mentioned earlier, we feed the canonical simulations with particle numbers 
within the slit pore obtained by using a cubic spline integration from our 
FMT-type theory results. Here, the simulations always recover the FMT-predicted bulk. 
Although the particle numbers obtained from FMT-PB and FMT-PB-MSA theory differ only 
slightly, corresponding simulations always fit the FMT-PB-MSA results better (not shown). 
Therefore, we choose the particle numbers 
predicted from FMT-PB-MSA theory as an input for our simulations. 
Figure~\ref{fig:conc_profiles} illustrates that FMT-PB-MSA theory excellently covers the details of 
the concentration profiles $\rho_{\pm}$ obtained by the simulations. Only in regions 
of high inhomogeneity, small deviations can be observed, e.g. in 
figure \ref{fig:conc_profiles}(i) in the midplane. 
Interestingly, figure \ref{fig:conc_profiles}(f) shows 
larger deviations between theory and simulations for the region around $z=0.9$nm than can be 
observed at higher potentials. 
However, it is remarkable that even at high potentials and in strong confinements FMT-PB-MSA is 
in very good agreement with simulations. 

\section{Charging and capacitance}

In the previous section we analyzed the potential and ion-concentration 
profiles for two pore widths of our model supercapacitor at a set of 
electrode potentials. Since we are interested in the capacitive performance of 
our model supercapacitor, we now study the relation between the surface charge 
density $e\sigma$ and the potential $\Psi$ on the electrode. This relation is 
naturally related to the differential capacitance per unit surface area 
\begin{equation}\label{eq:diff_cap}
C_{\rm diff}=\frac{\partial e\sigma}{\partial \Psi} . 
\end{equation}

\begin{figure}
\centering
\begin{picture}(8.0,16.1)
\put(0.0,5.3){\includegraphics[width=8.0cm]{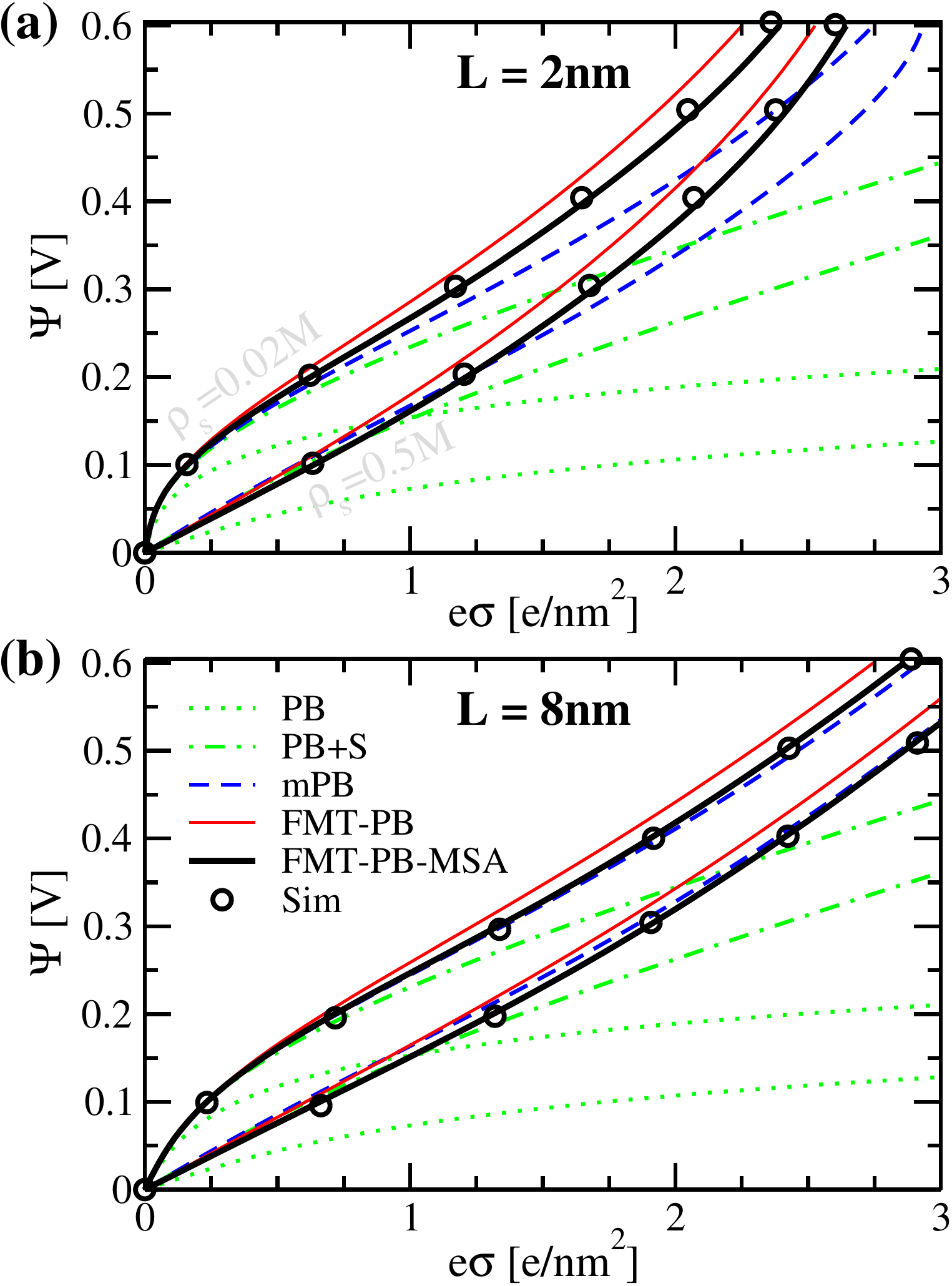}} 
\put(0.0,0.0){\includegraphics[width=8.0cm]{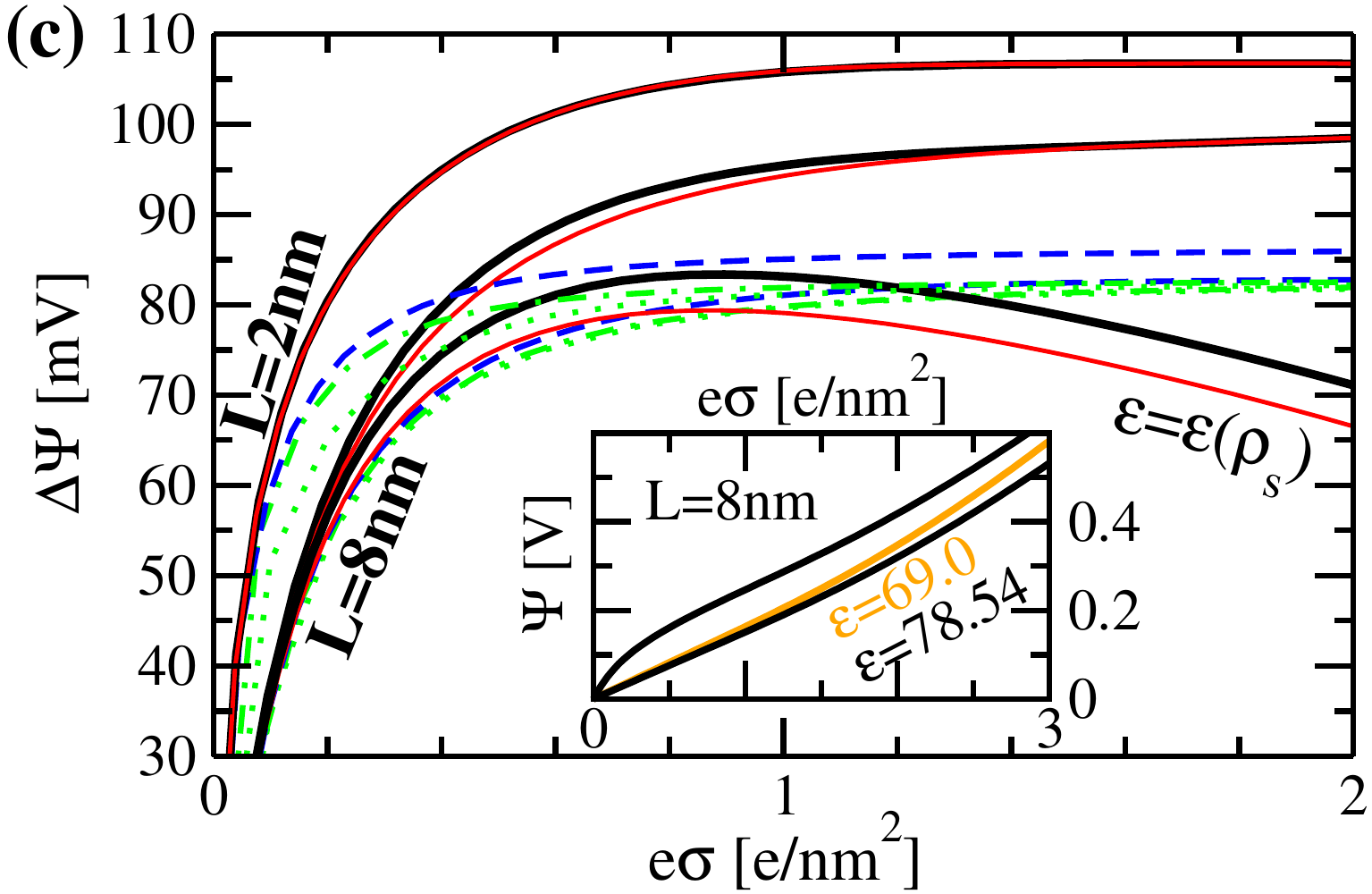}} 
\end{picture}
\caption{\label{fig:charge_potential_strokes} 
The charge-potential relations at two concentrations $\rho_{\rm s}=0.02$M and 
$\rho_{\rm s}=0.5$M shown for two pore sizes (a) $L=2$nm and (b) $L=8$nm. 
The potential rise $\Delta\Psi:=\Psi_{c=0.02\rm M}-\Psi_{c=0.5\rm M}$ 
between both {\it strokes} is shown in (c). In addition, we add data for the FMT-PB-MSA 
theory where we use a concentration-dependent $\varepsilon$ (see text) 
which depends on the bulk concentration $\rho_{\rm s}$ and is 
constant along each grand canonical stroke. }
\end{figure} 
In figures \ref{fig:charge_potential_strokes}(a) and (b) we plot the relation between 
the electrode potential and the electrode charge density at fixed 
bulk salt concentrations 
$\rho_{\rm s}=0.5$M and $\rho_{\rm s}=0.02$M, which  
correspond to sea and fresh river water, respectively. 
Since the electrode charge is limited by the maximum 
number of ions in a pore, the limit of close-packing in a slit-pore \cite{oguz_prl109_2012} 
sets the maximum electrode charge. Therefore, 
the surface charge density of a $2$nm pore 
can not exceed the maximum of approximately $3.9$e/nm$^{2}$. 
An example for such a densely packed system 
can be seen in figure \ref{fig:conc_profiles}(i). The corresponding 
charge-potential curve in figure \ref{fig:charge_potential_strokes}(a) is 
bound by this maximum charge density and indeed seems to converge towards it within 
the plotted region. 
In contrast, the PB and PB+S theories do not predict such a 
limiting charge, because they assume point-like ions. 
Thus, the electrode charge becomes unphysically high when potentials increase. 
A further comparison between figures \ref{fig:charge_potential_strokes}(a) and (b) 
shows that the electrode potential at a fixed electrode charge is higher at smaller 
pore size within all mentioned theories. 

In the next section we will discuss cyclic (dis)charging processes that follow
the charge-potential strokes at low and high salt concentrations. 
We will see that the amount of energy, which a blue engine can harvest from 
a concentration gradient, is related to the voltage rise between these 
strokes. This rise is shown in figure \ref{fig:charge_potential_strokes}(c) 
and converges towards a plateau for every choice of theory, provided that the 
dielectric constant is kept fixed. 
Since at low potentials the packing effects are not particularly pronounced, 
PB-type theories perform reasonably well in this regime. Deviations between the 
theories only start to appear at higher potentials. 
In general, the maximum 
value of the plateau seems to be related to the charge density $e\sigma$ at 
which the plateau is reached. Here, the more elaborate FMT-type theories 
reach the plateau at higher surface charges than the less elaborate ones. 
However, within all these theories the mentioned plateau is reached at higher 
charge densities $e\sigma$ when the pore size is increased. As a result, 
we find that the FMT-type theories reach the highest value for $\Delta\Psi$, followed 
by mPB and finally PB(+S) theory with the lowest value. 

The largest possible electrostatic 
potential rise is limited by the chemical potential difference between the involved solutes 
at low and high concentration. This can be seen from the limit where coions are 
fully excluded from the pore and counterions are most densely packed. 
In contact with a reservoir at either low or high concentration, the number of coions always 
stays the same and the chemical potential of the reservoir sets an offset for the 
electrostatic potential. For PB theory this 
limiting difference is well-known at $\Delta\Psi\approx83$mV 
for the salt concentrations used here, 
but it shifts, when other contributions than ideal gas are considered. For the additional 
steric contributions from the hard-sphere interactions in our FMT-type theories, the limit 
is higher at $\Delta\Psi\approx107$mV. 
Thus, the aforementioned corresponding waterfall \cite{norman_science186_1974} would be 
higher by a factor $107/83\approx 1.29$. 

Until now, we used a fixed dielectric constant of $\varepsilon=78.54$. 
However, measurements show that the bulk dielectric constant depends on temperature 
and concentration \cite{fernandez_jpcrd26_1997,sambriski_BiophysJ96_2009}. Keeping 
the temperature fixed at $T=298.15$K, this dependency reads 
$\varepsilon(\rho_{\rm s})=78.461 - 20.0154 \rho_{\rm s}/M+4.0415(\rho_{\rm s}/M)^2-0.54052(\rho_{\rm s}/M)^3$, 
spanning a range from $\varepsilon(0.02{\rm M})=78.06$ to $\varepsilon(0.5{\rm M})=69.4$. 
Taking this into account, the grand canonical charge-potential strokes change towards 
higher potentials (see inset of figure \ref{fig:charge_potential_strokes}(c)) 
and the potential rise $\Delta\Psi$ reaches a maximum at 
a finite charge density, decreasing again for higher charges. This behaviour is 
shown in figure \ref{fig:charge_potential_strokes}(c) within the FMT-PB-MSA 
theory. 


\begin{figure}
\centering
\includegraphics[width=8.5cm]{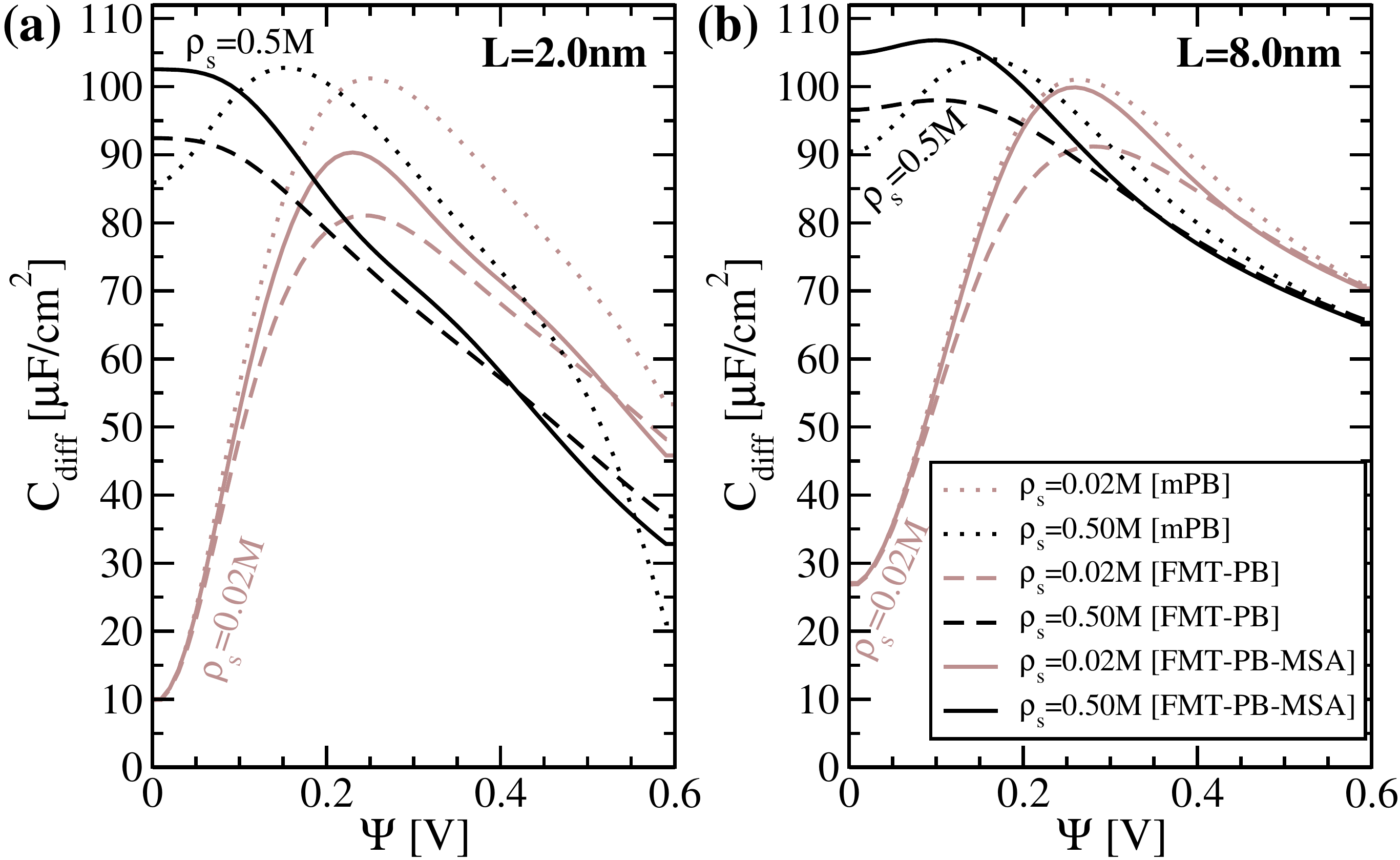}
\caption{\label{fig:diffcap}
Differential capacitances at two bulk concentrations $\rho_{\rm s}=0.02$M 
and $\rho_{\rm s}=0.5$M for two pore sizes (a) $L=2$nm and (b) $L=8$nm. }
\end{figure} 

The highly non-linear relation between charge and potential on the electrode 
is well captured by the differential capacitance as defined in equation~(\ref{eq:diff_cap}). 
It is known that in the limiting case of infinite electrode separation 
within PB theory $C_{\rm diff}$ 
increases with the electrode potential as $\sim\cosh(\phi/2)$, since there is no 
limiting maximal charge \cite{gouy1910constitution,Chapman1913}. For the PB+S theory, 
the differential capacitance reaches a plateau at high voltage, corresponding to 
the capacitance of the Stern layer. 
Within mPB theory at infinite plate separation, the differential 
capacitance was found analytically \cite{kornyshev2007double}. 
Contrary to the exponential increase of Gouy-Chapman, 
this solution decreases with the square root of the potential for large potentials. 
For low concentrations it has a local minimum 
at $\Psi=0$V, which turns into a global maximum at higher packing fractions. This 
transition from the so-called bell to camel shape occurs when the packing parameter 
within mPB theory becomes $\gamma=\tfrac{1}{3}$ 
and both camel and bell shape have been found 
experimentally \cite{islam_jpcc112_2008,cannes_jpcc117_2013}.
%
%
%
%
%
%
Obviously, in the confining geometry of our model supercapacitor, both mPB and FMT-type 
theories reproduce decaying tails of $C_{\rm diff}$ for increasing electrode potentials, as shown 
in figure \ref{fig:diffcap}. Interestingly, at high salt concentration the FMT-type 
theories already passed to a bell shape in the micropores in figure \ref{fig:diffcap}(a), 
whereas mPB theory still predicts a camel shape. 
This finding can also be seen from figure~\ref{fig:charge_potential_strokes}(a), where the 
stroke corresponding to mPB theory at high concentration shows a convex curvature at low 
potential instead of a concave one when compared to the FMT-type results, which helps to 
reach the plateau in figure~\ref{fig:charge_potential_strokes}(c) at a 
small charge density $e\sigma$. 
Moreover, figure~\ref{fig:conc_profiles} shows that within small pores the concentration 
profiles in the midplane are already affected at an electrode potential $\Psi=0$V, which could 
explain the bell shape in our FMT-type results.

\section{Cyclic processes}

Our model supercapacitor can be charged (discharged) by connecting it to (using it as) 
an external power source. Thereby, the electric double layer is built up at (removed from) 
the porous electrodes of the system, storing (releasing) energy. 
If (dis)charging happens sufficiently slowly, the dynamical processes can be assumed to 
be quasi-static. Thereby, the (dis)charging time should be substantially larger than the 
$RC$-time, 
such that the system always has enough time to 
equilibrate and therefore stays in equilibrium. 
Even though real systems typically work on shorter time scales out of equilibrium 
we investigate the universal quasi-static behaviour as a starting point. 
If, now, the electrodes are in contact with a reservoir at fixed chemical potential 
while they are (dis)charged, the (dis)charging process is described by the 
grand canonical (dis)charging strokes 
which are shown in figure~\ref{fig:charge_potential_strokes}. 

\begin{figure*}
\centering
\centering
\includegraphics[width=15.0cm]{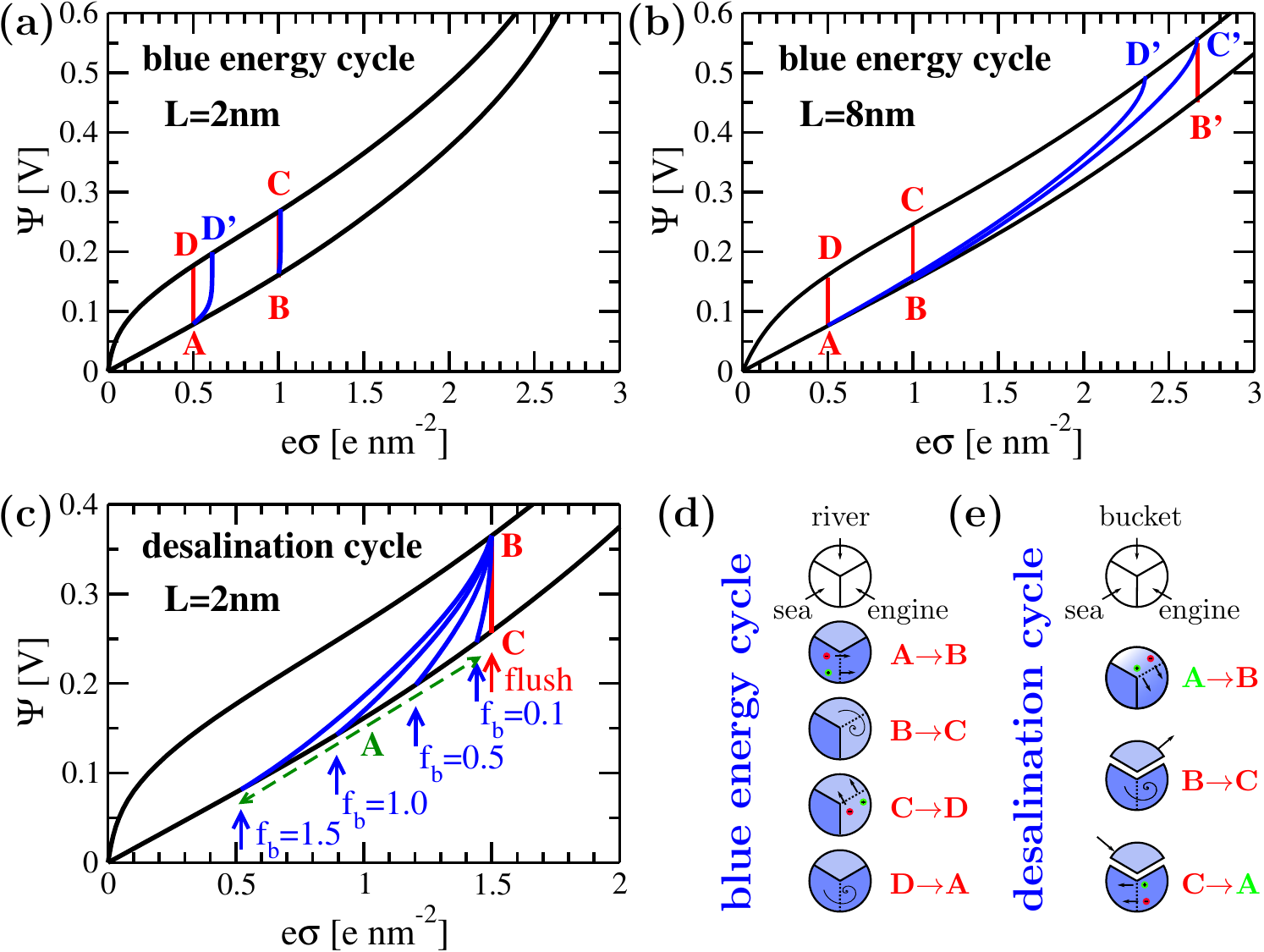}
\caption{\label{fig:cycles}
Blue-energy and desalination cycles in between 
charge-potential strokes at two concentrations $\rho_{\rm s}=0.02$M 
and $\rho_{\rm s}=0.5$M for two pore sizes $L=2$nm (left-hand side) and 
$L=8$nm (right-hand side). More detail about the cycles are given in the 
text and in the process diagrams in (d) and (e).}
\end{figure*}
\begin{figure}
\centering
\includegraphics[width=8.2cm]{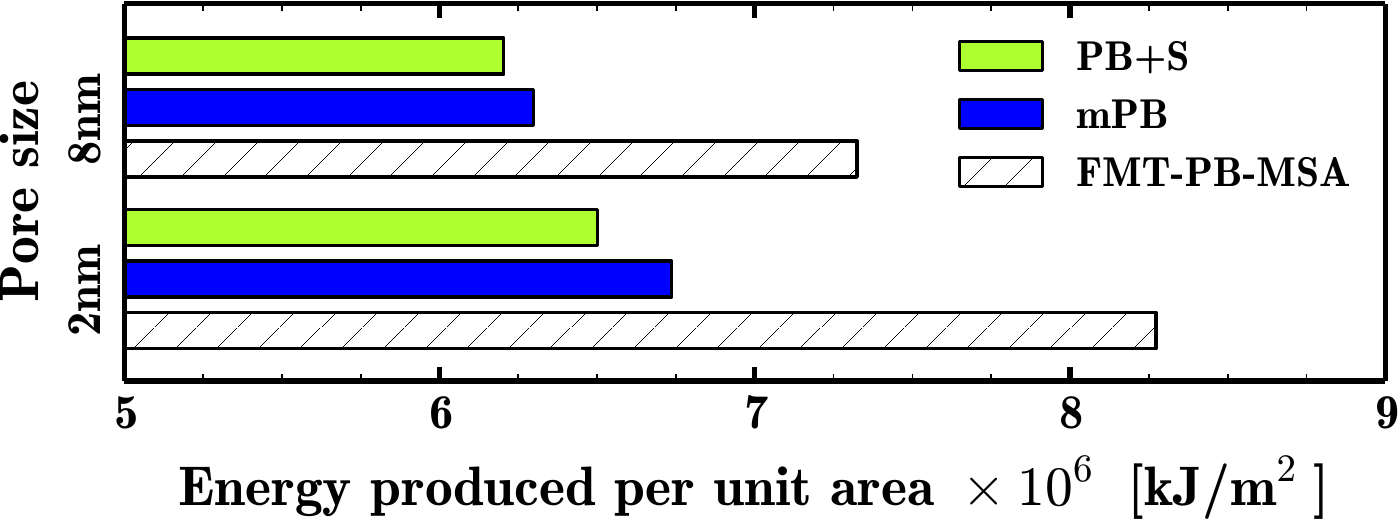}
\caption{\label{fig:blue}
Energy harvested per unit surface area of the electrode material during one 
blue-energy cycle ABCDA as shown in figures~\ref{fig:cycles}(a) and (b). 
We compare PB+S, mPB and FMT-PB-MSA theory for both pore size $L=2$nm and $L=8$nm. }
\end{figure}

 
In the previous section, we have seen that these (dis)charging strokes depend 
on the salt concentration within the reservoir, to which the electrodes are connected. 
To exploit this dependency, Brogioli recently proposed a four-stroke 
charging-desalination-discharging-resalination cycle 
to harvest a net amount of energy \cite{brogioli2009extracting}. 
This process extracts the free energy which is released 
during the selective mixing of water at two different concentrations. 
Such blue-energy cycles ABCDA are shown in figures~\ref{fig:cycles}(a) and (b) and are 
explained schematically in figure~\ref{fig:cycles}(d). The cycle starts with a grand 
canonical charging step AB, while the engine is connected to a reservoir at high ion 
concentration. This step is followed by a flushing step, where the water at high 
ion concentration is replaced by water at low ion concentration, which 
rises the electrode potential while the charge on the electrode remains the same. 
Now, a grand canonical discharging step follows and, finally, the engine is brought 
back to the initial state by flushing with sea water. 
In fact, the total work 
\begin{equation}\label{eq:work}
W=-\oint \Psi dQ
\end{equation}
delivered in such a cycle is positive, because the charging steps in sea water 
take place at a lower potential than the discharging steps in river water. 
In figure~\ref{fig:blue} we present the corresponding energy harvested during such 
blue-energy cycles ABCDA for the two pores of size $L=2$nm and $L=8$nm within 
three of our theories. In agreement with figure~\ref{fig:charge_potential_strokes} 
we find that the energy per unit surface area of the electrode is larger for 
smaller pores. Furthermore, 
it is obvious from figures~\ref{fig:cycles}(a) and (b) that the area enclosed by the cycle ABCDA 
increases with the potential rise 
$\Delta\Psi$ between the strokes at low and high concentration. This quantity, 
previously discussed and depicted in figure~\ref{fig:charge_potential_strokes}(c), 
was shown to be largest for the elaborate FMT-type theories, which is also depicted in 
figure~\ref{fig:blue}. 
Since this enhanced work output is a model-dependent finding, it does not 
correspond to a higher work output in an experimental setting, only to a more 
accurate description of the physics at hand. 
For example, along with figure~\ref{fig:charge_potential_strokes}(c) we also 
discussed the consequences of using a dielectric constant that depends on the 
concentration, which results in an overall lower work output than a corresponding 
cycle with a fixed dielectric constant would have. 
Nevertheless, the mutual deviation between the theories shows that packing effects are 
important in the context of cyclic processes such as discussed in this work 
and should be taken into account.

In order to accomplish an actual increase in the work output, 
we distinguish two possibilities. 
First, the electrostatic potential not only depends on the chemical potential,
but also on a set of variables like temperature, valence or pore 
size. In principle, they can be tuned to make the potential rise between the 
two involved concentrations and the corresponding work output as large as possible. 
For instance, a threefold increase of the work output is possible, when instead of 
sea and river water at equal temperature, high-temperature (e.g. waste-heated) 
river water is used during the resalination step \cite{Janssen:2014aa}. This 
effect has also been observed experimentally \cite{ahualli_est48_2014}. 
Furthermore, varying water temperature can also enhance the work output 
in a membrane-based setup \cite{sales_estl1_2014}. 

Second, the energy dissipated during operation can be made as small as possible by 
choosing an optimal sequence of ensembles for the (dis)charging steps. Similar 
to Carnot's optimization of the Stirling cycle, the blue-energy cycle can be optimized 
with respect to the harvested energy per ion \cite{boon2011blue}. By mapping the 
intensive and extensive thermodynamic variables of the blue engine to those of a heat 
engine system, which operates on a thermal instead of a concentration gradient, 
it was shown that the Brogioli cycle is the CAPMIX analogue of a Stirling 
cycle \cite{Roij:2012aa}. Accordingly, an optimal blue engine would have a rectangular 
path in $\mu N$-representation of chemical potential $\mu$ and ion number $N$, 
that is, two grand-canonical steps at fixed $\mu$ and two (new) canonical steps 
(BC' and D'A) at fixed $N$. Then, each ion transported between the two reservoirs is 
able to donate its full $\Delta\mu$ to the harvested energy without spoiling energy 
in the irreversible flushing steps (BC and DA). 

The order of the canonical and grand-canonical strokes is fixed by the 
thermodynamic relation $C_{\mu}>C_{N}$ between 
the differential capacitances at fixed chemical potential $\mu$ or particle number $N$, 
respectively \cite{Roij:2012aa}. 
Furthermore, from figures~\ref{fig:cycles}(a) and (b) it becomes obvious that the slope 
of the canonical strokes in the $\Psi\sigma$-representation highly depends on packing properties and, 
accordingly, changes when varying, for example, the particle density 
or the pore size. 
Thereby, a steep slope hints at a situation close to a limiting 
charge on the electrode due to close packing of ions in the pore, as discussed 
in the previous sections. 
Accordingly, the canonical strokes in figures~\ref{fig:cycles}(a) and (b) become steeper 
in the narrow micropore of $L=2$nm than they are in the mesopore of $L=8$nm where 
the center of the pore still can act as a reservoir for counterions. 
For very dense packings, the canonical strokes become almost vertical in the 
$\Psi\sigma$-representation, as we show in figure~\ref{fig:cycles}(a) where 
the point C' is not shown since it is almost at the same position ($\sigma=1.01$nm$^{-2}$)
as the point C ($\sigma=1.00$nm$^{-2}$). 

As is obvious from figure~\ref{fig:cycles}(b), the Carnot-like 'optimal' (with respect to the harvested 
energy per ion) cycle ABC'D'A has a smaller area 
than the cycle AB'C'DA. Both cycles are performed in the same potential 
range and for a pore volume of $V_{\rm e}=2 V_{\rm el}=1$L within both 
electrodes, the work output of the engine is $W\approx 1.3$kJ and 
$W\approx 8.4$kJ, respectively. 
Thus, a discrepancy exists between an optimal cycle in the 
sense of total harvested energy per cycle and harvested energy per ion. 
Which of the two options is more desirable depends on the availability of the required resources, 
in particular that of fresh water.\\
In addition, a well-chosen combination 
of canonical and grand canonical strokes is able to tune the number of ions that 
is passed through the engine during operation. 
For example, the energy output for the cycles ABC'D'A and 
AB'C'DA in figure~\ref{fig:cycles}(a) for an engine volume of $1$L is 
$W\approx 6.8$kJ and $W\approx 8.3$kJ, respectively. 
Interestingly, the energy output of the cycle AB'C'DA seems to be similar for both 
pore sizes we have analyzed. This also holds for the PB+S and mPB theory (not shown), 
where the energy harvested during this cycle is $W\approx 7.8$kJ and $W\approx 7.2$kJ (PB+S) 
and $W\approx 7.5$kJ and $W\approx 7.3$kJ (mPB) for the two pores of $L=2$nm and $L=8$nm. 

Similar to its heat engine counterpart, the blue engine can also be run 
in reverse for desalinating water. An example of the corresponding three-stroke cycle ABCA 
is shown in figure~\ref{fig:cycles}(c) and explained schematically in 
figure~\ref{fig:cycles}(e). Starting in an initial state A, we connect 
the capacitor of pore volume $V_{\rm e}$ to a bucket of volume $V_{\rm b}$ which is to
 be desalinated. The capacitor is charged canonically, capturing the ions 
into the capacitor pores and desalinating the bucket. After reaching the low concentration 
in point B, we disconnect the bucket to harvest the fresh water, and flush the capacitor 
with water at high ion concentration to point C. We then simultaneously refill 
the bucket with seawater, and discharge the capacitor in contact with
the sea reservoir until the initial charging state A is reached. Here the bucket and capacitor are
reconnected so that the cycle can start again. 

Keeping point B fixed, the location of point A in 
figure~\ref{fig:cycles}(c) depends on the chosen fraction $f_{\rm b}=V_{\rm b}/V_{\rm e}$ 
between the bucket and engine volumes, because the size of the connected bucket 
also sets the number of ions that must be adsorbed into the capacitor. 
For this reason, the slope of the canonical strokes AB depends on the fraction $f_{\rm b}$.
Note that this quantity is varied here by changing the bucket size, keeping the pore width and volume fixed.

\begin{figure}
\centering
\includegraphics[width=8.2cm]{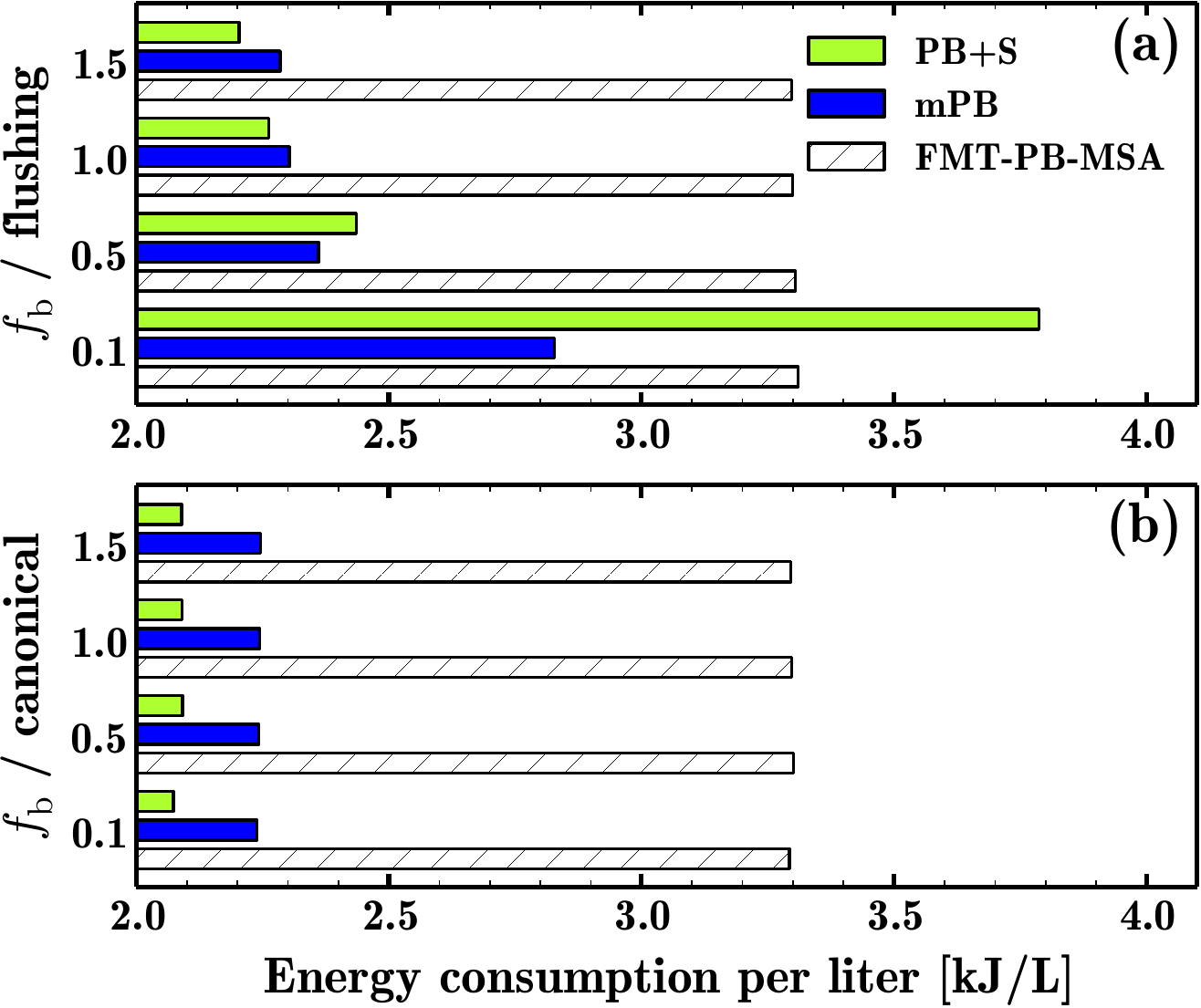}
\caption{\label{fig:desal2}
Energy consumption per liter of desalinated water for desalination cycles as 
shown in figure~\ref{fig:cycles}(c), where point B lays at $\sigma=1.5$nm$^{-2}$. 
We compare PB+S, mPB and FMT-PB-MSA theory for four ratios 
$f_{\rm b}=V_{\rm b}/V_{\rm e}$ between bucket and engine 
volumes for (a) a flushing step BC and (b) a canonical step BC', respectively. }
\end{figure} 

Unlike a blue-energy cycle, the work in equation~(\ref{eq:work}) picks up 
an extra minus sign in a desalination cycle because the latter runs clockwise 
instead of anti-clockwise in its charge-potential representation. Consequently, 
desalination costs energy. 

Similar to the previously discussed blue-energy cycle, to minimize energy losses
the dissipative flushing step BC can be replaced by a reversible canonical step BC'. 
Though we will discuss these cycles in the following, 
we do not show point C' in figure~\ref{fig:cycles}(c) because it would almost coincide 
 with C. Note that for larger pore sizes the points C and C' would 
separate more, as shown in figures~\ref{fig:cycles}(a) and (b) for a blue-energy cycle. 
The interior of larger pores namely acts more like a reservoir, such that it 
requires a more significant ion-release from the EDL to resalinate a larger pore.
In figure~\ref{fig:desal2} we compare the energy consumption during desalination 
cycles as shown in figure~\ref{fig:cycles}(c), for different bucket fractions 
$f_{\rm b}$ within PB+S, mPB and FMT-PB-MSA theory for a flushing step BC and 
a canonical step BC', respectively. 

For the flushing-type desalination cycle, we find that FMT-type calculations predict the largest 
amount of consumed energy in most cases when compared to PB+S and mPB theory. 
This was to be expected on the basis of figure~\ref{fig:charge_potential_strokes}(c) 
which shows that potential steps with respect to different concentrations are 
largest in FMT-type theories. Furthermore, in figure~\ref{fig:desal2}(a) PB+S and 
mPB theory predict an increasing amount of required energy to desalinate one liter of 
water for decreasing fractions $f_{\rm b}$ and, accordingly, bucket volumes $V_{\rm b}$, 
while FMT-PB-MSA theory predicts quite constant values of invested energy. This 
can be understood by analyzing the cycle ABCA in the limit of vanishing bucket 
volume $V_{\rm b}$ and fraction $f_{\rm b}\to 0$. In this case of vanishing bucket size, the point A reaches 
the point C' and the desalination step becomes equivalent to a canonical step without 
desalination bucket. In other words, the work of the cycle ABCA becomes equivalent 
to the work that corresponds to the enclosed area C'BCC', which remains finite. 
For this reason, the energy per volume of desalinated water diverges in this limit. 
However, in FMT-type theories the enclosed area C'BCC' is much smaller 
than in PB+S and mPB theory due to the previously discussed 
packing effects and, therefore, the invested work seems to stay unaffected from 
the bucket size within the explored range $[0.1,1.5]$ of fraction $f_{\rm b}$. 
Nevertheless, it will diverge if $f_{\rm b}$ becomes small enough. 

If the flushing step BC is replaced by a canonical step BC', 
the enclosed area C'BCC' reduces to a single line which does not 
enclose any area. 
Therefore, the work per cycle vanishes for vanishing 
bucket volume $V_{\rm b}$ and, for this reason, no divergences should appear, as 
indeed observed in figure~\ref{fig:desal2}(b). Moreover, the desalination energy per liter water 
is constant for all fractions $f_{\rm b}$ within all mentioned theories, because 
the corresponding cycles are optimal with respect to the energy per ion. 
Accordingly, the energy per liter of desalinated water is constant. 

For larger pores of $L=8$nm, during a desalination stroke, the total capacitor
potential would almost reach the limit of $1.2$V, where electrochemical processes would start. 
The meso- (and macropores) therefore lower the pore-averaged capacitance, 
and operating real porous carbon electrodes between sea and river concentrations 
could be problematic. To avoid such problems, it might be interesting to operate 
a series of blue engine capacitors with intermediate reservoirs at intermediate 
concentrations instead of using only one engine over the whole concentration gradient. 
Here, balancing the flushing and canonical steps could allow for tuning the intermediate 
concentrations. We have not pursued this next optimization step in detail yet. 

\section{Discussion and conclusions}


In this article, we studied cyclic processes involving 
supercapacitors immersed in water at low and high salt concentration 
for the purpose of energy harvesting and desalination. 
In this context, we investigated ions within the restricted primitive 
model in the micro- and mesopores of the supercapacitor electrodes 
at different electrostatic potentials. We compared a set of 
available theories, ranging from simple Poisson-Boltzmann 
up to sophisticated density functional theories. To distinguish between 
these theories and define their applicable parameter settings, we also 
performed molecular dynamics simulations. We found best agreement with these 
simulations for a fundamental measure density functional theory with mean-field 
Coulombic ion interactions, taking excess correlations beyond the pure Coulomb 
and pure hard interactions into account \cite{mieryteran_jcp92_1990}. 
In agreement with earlier 
studies \cite{mieryteran_jcp92_1990,tang_MolPhys71_1990,yu_cjce12_2004,yu_jcp_2004,frydel_jcp137_2012}, 
we found that ion sizes cannot be neglected if a qualitative or, even more, 
a quantitative description of systems similar to those we studied is of 
interest. For mesopores, dilute ion concentrations or low potentials, 
simple theories like modified Poisson-Boltzmann 
theory \cite{kornyshev2007double} are still applicable. 
However, we show that in the micropores, where the hydrated 
ion size becomes of the same order of magnitude as the pore width,
packing effects {\it do} become important, and the mean field theories 
do not agree with simulations. For such settings, 
correlations between the ions must be taken into account in more 
elaborate theories like our density functional approach. 

In the context of cyclic processes, our density functional approach predicts 
a significant larger amount of energy that can be harvested from 
blue engines than the more simple theories in our study. Relatedly, it 
predicts a larger consumption of energy for the desalination of water. 
Furthermore, we found that the (theoretical) description of the solvent 
has a high impact on these predictions. 
Here we found that taking into account the 
dependency of the dielectric constant on the salt concentration corrects 
the amount of available energy due to a concentration gradient downwards. 
Moreover, this treatment of the solvent predicts a potential range 
for the optimal operation of a blue engine around an electrode potential 
$\Psi\approx 0.2$V. Consequently, a desalination device could be optimized 
by shifting this operational range to lower or higher potentials. 


Within further research our theoretical description could be improved. 
While we described the solvent by a uniform dielectric constant, an explicit 
description of the water molecules could lead to a more accurate picture of the 
electric double layer \cite{biben_pre57_1998,henderson_jpcb116_2012}. 
Moreover, for dense packings, an explicit description of the 
hydration shell becomes important \cite{levy_prl108_2012,bankura_jcp138_2013} 
to allow for the dehydration of ions near walls or in the smallest 
pores \cite{ohkubo_jacs124_2002,merlet2012molecular} and for similar effects, where the 
polarizability of the water molecules plays a role. 
Furthermore, we described the rich geometry of the porous carbon electrode 
by a relatively simple parallel-plate capacitor model, neglecting effects 
due to the pore curvature \cite{huang_cej14_2008} or roughness of the 
electrode surface \cite{merlet_jpcl4_2013}. 
In addition, 
carbon electrodes show metallic behaviour and accounting for image 
charge effects can modify the salt adsorption at low voltages \cite{biesheuvel_jsse18_2014}. 
Nevertheless, our findings from the parallel plate model might have 
interesting applications for the conversion between mechanical and electric energy \cite{moon_nc4_2013}. 
Finally, dynamics can be taken into account to address the possibly slow ion transport 
through the pore network of the carbon electrode. This could provide the 
power performance of a blue engine which might highly deviate from the 
quasi-static calculations, which only address the energy of a cycle. However, 
recent simulation results show that agreement between the quasi-static calculations 
and dynamical studies is better than might be 
expected \cite{pean_nano8_2014}. 

Further research could also study the implications of our findings on so-called 
flow electrodes \cite{heeacho2013desalination,porada2014carbon}. These devices 
provide an interesting alternative to the solid porous carbon electrodes used in 
conventional desalination devices. They consist of micron size particles of nano 
porous carbon in a suspension which allows to control the salt uptake by varying 
the feed velocity of the carbon slurry. 

In a more general context, our findings also are suitable for studying the storage of energy 
in ionic-liquid supercapacitors. In such systems, the screening length is much 
shorter than in an aqueous solution. In addition, ionic liquids have a larger 
potential window in comparison to water. For this reason, higher potentials 
can be applied and ionic packing effects become even more important. Nevertheless, 
even very high packings of spheres are well described within fundamental measure 
density functional theory. Moreover, numeric calculations within this theoretical 
framework are much faster than computer simulations. Thus, the studied approach 
represents a promising starting point for more complex models or to 
derive from it more advanced theoretical descriptions for future research.

\section*{Acknowledgements}  

We thank D.~Brogioli and Y.~Levin for fruitful discussions. 
This work is part of the D-ITP consortium, a program of the 
Netherlands Organisation for Scientific Research (NWO) that is funded by 
the Dutch Ministry of Education, Culture and Science (OCW). 
We also acknowledge financial support from a NWO-VICI grant.

\providecommand{\newblock}{}

\end{document}

%% file: setup.pdf_tex
\begingroup%
  \makeatletter%
  \providecommand\color[2][]{%
    \errmessage{(Inkscape) Color is used for the text in Inkscape, but the package 'color.sty' is not loaded}%
    \renewcommand\color[2][]{}%
  }%
  \providecommand\transparent[1]{%
    \errmessage{(Inkscape) Transparency is used (non-zero) for the text in Inkscape, but the package 'transparent.sty' is not loaded}%
    \renewcommand\transparent[1]{}%
  }%
  \providecommand\rotatebox[2]{#2}%
  \ifx\svgwidth\undefined%
    \setlength{\unitlength}{209.99605371bp}%
    \ifx\svgscale\undefined%
      \relax%
    \else%
      \setlength{\unitlength}{\unitlength * \real{\svgscale}}%
    \fi%
  \else%
    \setlength{\unitlength}{\svgwidth}%
  \fi%
  \global\let\svgwidth\undefined%
  \global\let\svgscale\undefined%
  \makeatother%
  \begin{picture}(1,0.75392163)%
    \put(0,0){\includegraphics[width=\unitlength]{setup.pdf}}%
    \put(0.74750663,0.54351164){\color[rgb]{0,0,0}\makebox(0,0)[lt]{\begin{minipage}{0.28054991\unitlength}\raggedright $\Psi_{+}$\end{minipage}}}%
    \put(0.57985989,0.41743782){\color[rgb]{0,0,0}\makebox(0,0)[lt]{\begin{minipage}{0.24662522\unitlength}\raggedright $L\sim 2$ nm\\ \end{minipage}}}%
    \put(0.74750663,0.72213046){\color[rgb]{0,0,0}\makebox(0,0)[lt]{\begin{minipage}{0.28054991\unitlength}\raggedright $\Psi_{-}$\end{minipage}}}%
    \put(0.06766538,0.61549914){\color[rgb]{0,0,0}\makebox(0,0)[lt]{\begin{minipage}{0.46667544\unitlength}\raggedright $\rho_s$\end{minipage}}}%
    \put(0.04713793,0.14015969){\color[rgb]{0,0,0}\makebox(0,0)[lt]{\begin{minipage}{0.24662522\unitlength}\raggedright $\sim50$ nm\end{minipage}}}%
  \end{picture}%
\endgroup%